\documentclass[sigconf, nonacm]{acmart}

\AtBeginDocument{%
  \providecommand\BibTeX{{%
    \normalfont B\kern-0.5em{\scshape i\kern-0.25em b}\kern-0.8em\TeX}}}

\usepackage[]{graphicx}
\usepackage{booktabs}
\usepackage{amsmath}
\usepackage{algorithmic}
\usepackage{array}
\usepackage{comment}
\usepackage{mdwmath}
\usepackage{mdwtab}
\usepackage{eqparbox}
\usepackage{multirow}
\usepackage[tight,footnotesize]{subfigure}
\DeclareGraphicsExtensions{.pdf,.png,.jpg}
\usepackage{siunitx}

\usepackage{multirow}
\usepackage{multicol}
\usepackage{color}
\usepackage{xspace}
\renewcommand{\vec}[1]{\mathbf{#1}}

\usepackage[tight]{subfigure}
\usepackage{array}
\newcolumntype{P}[1]{>{\centering\arraybackslash}p{#1}}

\copyrightyear{2019} 
\acmYear{2019} 
\setcopyright{usgovmixed}
\acmConference[KDD '19]{The 25th ACM SIGKDD Conference on Knowledge Discovery and Data Mining}{August 4--8, 2019}{Anchorage, AK, USA}
\acmPrice{15.00}
\acmDOI{10.1145/3292500.3330703}
\acmISBN{978-1-4503-6201-6/19/08}

\begin{document}

\title[IRNet]{Deep Residual Learning on Vector Inputs 
to Predict Stability of Materials from Crystal Structures
}


\title{IRNet: A General Purpose Deep Residual Regression Framework for Materials Discovery
}

\author{Dipendra Jha}
\affiliation{%
  \institution{Northwestern University}
  \city{Evanston}
  \state{Illinois}
  \country{USA}
}
\email{dipendra.jha@eecs.northwestern.edu}

\author{Logan Ward}
\affiliation{%
  \institution{U.\ Chicago and Argonne National Lab}
  \city{Chicago}
  \state{Illinois}
  \postcode{USA}
}
\email{loganw@uchicago.edu}

\author{Zijiang Yang}
\affiliation{%
  \institution{Northwestern University}
  \city{Evanston}
  \state{Illinois}
  \country{USA}
}
\email{zyz293@eecs.northwestern.edu}

\author{Christopher Wolverton}
\affiliation{%
  \institution{Northwestern University}
  \city{Evanston}
  \state{Illinois}
  \country{USA}
}
\email{c-wolverton@northwestern.edu}

\author{Ian Foster}
\affiliation{%
  \institution{U.\ Chicago and Argonne National Lab}
  \city{Chicago}
  \state{Illinois}
  \country{USA}
}
\email{foster@uchicago.edu}

\author{Wei-keng Liao}
\affiliation{%
  \institution{Northwestern University}
  \city{Evanston}
  \state{Illinois}
  \country{USA}
}
\email{wkliao@eecs.northwestern.edu}

\author{Alok Choudhary}
\affiliation{%
  \institution{Northwestern University}
  \city{Evanston}
  \state{Illinois}
  \country{USA}
}
\email{choudhar@eecs.northwestern.edu}

\author{Ankit Agrawal}
\affiliation{%
  \institution{Northwestern University}
  \city{Evanston}
  \state{Illinois}
  \country{USA}
}
\email{ankitag@eecs.northwestern.edu}

\renewcommand{\shortauthors}{D. Jha et al.}

\begin{abstract}
Materials discovery is crucial for making scientific advances in many domains. Collections of data from experiments and first-principle computations have spurred interest in applying machine learning methods to create predictive models capable of mapping from composition and crystal structures to materials properties. Generally, these are regression problems with the input being a 1D vector composed of numerical attributes representing the material composition and/or crystal structure.
While neural networks consisting of fully connected layers have been applied to such problems, 
their performance often suffers from the vanishing gradient problem when network depth is increased.
Hence, predictive modeling for such tasks has been mainly limited to traditional machine learning techniques such as Random Forest.
In this paper, we study and propose design principles for building deep regression networks composed of fully connected layers with numerical vectors as input. We introduce a novel deep regression network with individual residual learning, IRNet, that places shortcut connections after each layer so that each layer learns the residual mapping between its output and input. 
We use the problem of learning properties of inorganic materials from numerical attributes derived from material composition and/or crystal structure to compare IRNet's performance against that of other machine learning techniques.
Using multiple datasets from the Open Quantum Materials Database (OQMD) and Materials Project for training and evaluation,
we show that IRNet provides significantly better prediction performance than the state-of-the-art machine learning approaches currently used by domain scientists.
We also show that IRNet's use of individual residual learning leads to better convergence during the training phase than when shortcut connections are between multi-layer stacks while maintaining the same number of parameters. 
\end{abstract}




\maketitle

\section{Introduction}
\label{sec:intro}
Materials discovery plays an important role in many domains of science and engineering~\cite{1367-2630-20-4-043010,qin2017mechanics}.
The slow pace of development and deployment of new/improved materials is
a major bottleneck in the innovation cycles of emerging technologies~\cite{kalidindi2015data}.
Collection of large scale datasets through experiments and first-principle computations such as high throughput density functional theory (DFT) calculations~\cite{curtarolo2013high, 
jain2013the, kirklin2015open
} and the emergence of integrated data collections and registries~\cite{blaiszik2016materials,dima2016informatics}
have spurred the interest of materials scientists in applying machine learning (ML) models to understand materials and predict their properties
~\cite{xue2016accelerated,botu2015adaptive,
ward2016general,faber2016machine,
ramprasad2017npjreview,liu2015predictive,
seko2017representation, 
pyzer2015learning,montavon2013machine}, leading to the novel paradigm of materials informatics~\cite{agrawal2016perspective,
 rajan2015materials,ward2016atomistic,
ramprasad2017npjreview}.
Such interests have been supported by government initiatives such as the Materials Genome Initiative (MGI)~\cite{mgi}.

Predictive modeling tasks in materials science are generally regression problems where we need to predict materials properties from an input vector composed of numerical features derived from their composition and/or crystal structures by incorporating domain knowledge~\cite{xue2016accelerated,botu2015adaptive,ward2016general,faber2016machine, ramprasad2017npjreview,seko2017representation, jha2018elemnet}.
Since the model input contains vector of independent features, the neural network models used for such tasks are composed of fully connected layers.
Vanishing gradient  and performance degradation issues that arise when using deeper architectures have caused
the neural network architectures used for such prediction modeling to be limited in their depth~\cite{montavon2013machine, pyzer2015learning, paul2018chemixnet, jha2018elemnet, zhou2018learning}.
For instance, Montavon et al.~\cite{montavon2013machine} trained a four-layer network on a database of around 7000 organic compounds to predict multiple electronic ground-state and excited-state properties.
In the Harvard Energy Clean Project, 
Pyzer-Knapp et al.~\cite{pyzer2015learning} used a three-layer network for predicting power conversion efficiency of organic photo-voltaic materials.
Zhou et al.~\cite{zhou2018learning} used a fully connected network with single hidden layer to predict formation energy from high-dimensional vectors learned using Atom2Vec. 
ElemNet~\cite{jha2018elemnet} used a 17-layered architecture to learn formation energy from elemental composition, but experienced performance degradation beyond that depth.
Hence, domain scientists have mainly used traditional ML techniques such as Random Forest, Kernel Ridge Regression, Lasso, and Support Vector Machines for materials prediction tasks~\cite{ghiringhelli2015big, faber2015crystal, meredig2014combinatorial, ward2016general}.

Recently, several projects have used domain knowledge-based model engineering within a deep learning
context for predictive modeling in materials science~\cite{schutt2017schnet,jha2018extracting,goh2017smiles2vec}.
Deep learning was used for directly predicting the crystal orientations of polycrystalline materials from their electron back-scatter diffraction patterns~\cite{jha2018extracting}.
SchNet~\cite{schutt2017schnet} used continuous filter convolutional layers to model quantum interactions in molecules for the total energy and interatomic forces that follows fundamental quantum chemical principles.
Boomsma and Frellsen~\cite{boomsma2017spherical} introduced the idea of spherical convolution in the context of molecular modelling, by considering structural environments within proteins.
Smiles2Vec~\cite{goh2017smiles2vec} and CheMixNet~\cite{paul2018chemixnet} have applied deep learning methods to learn molecular properties from the molecular structures of organic materials.

Our goal here is to design a general purpose deep regression network for predicting the properties of inorganic materials from their compositions and/or crystal structures, without using any domain knowledge-based model engineering.
We introduce the idea of residual learning to deep regression networks composed of fully connected layers.
In a fully connected network, the number of parameters is directly proportional to the product of the number of inputs and the number of output units.
Several works have dealt with the performance degradation issue due to vanishing or exploding gradients for other types of data mining problems~\cite{srivastava2015training, he2016deep, huang2017densely}.
Srivastava et al.~\cite{srivastava2015training} introduced an LSTM-inspired adaptive gating mechanism that allowed  information to flow across layers without attenuation; the gating mechanism required more model parameters.
They designed highway networks composed of up to 100 layers that could be optimized.
A highway network~\cite{srivastava2015training} uses gated connections, which double the number of parameters in a fully connected network.
In a DenseNet~\cite{huang2017densely}, all previous inputs are combined before being fed into the current layer. 
For a fully connected network, this approach results in a tremendous increase in the number of model parameters,
a particular problem when working with limited GPU memory.
He et al.~\cite{he2016deep} introduced the idea of residual learning,
in which a stack of layers learns the residual mapping between the output and input;
they built deep CNN models composed of 152 layers for image classification problem.
Since the input is added to the residual output, the number of required parameters for residual learning was lower than that in Srivastava et al.~\cite{srivastava2015training}.
This technique has been used in several CNN and LSTM architectures, with shortcut connections being placed after a stack of multiple CNN or LSTM layers to build deeper networks for better performance~\cite{szegedy2017inception, wang2016recurrent,huang2017improved
}.
For a fully connected network, an elegant approach is to use the residual mapping approach used in ResNet~\cite{he2016deep}.
However, although residual learning has been widely used in classification networks, no previous work leverages residual learning for building deep regression networks composed of fully connected layers for numerical vector inputs. 

In this paper,
we study and propose design principles for building deep residual regression networks
composed of fully connected layers for data mining problems with numerical vectors as inputs.
We introduce a novel deep regression network architecture with individual residual learning (IRNet), 
in which shortcut connections are placed after each layer such that each layer learns only the residual mapping between its output and input vectors.
We compare IRNet against two baseline deep regression networks:
and a stacked residual network (SRNet) with shortcut connections after stack of multiple layers. 
We focus on the \emph{design problem} of learning the formation enthalpy of inorganic materials from an input vector composed of 126 features representing their crystal structure, 
and another 145 composition-based physical attributes from the OQMD-SC dataset.
OQMD-SC contains \num{435582} materials with their composition and crystal structure from the Open Quantum Materials Database (OQMD)~\cite{kirklin2015open}.

Our proposed 48-layered IRNet achieves significantly better performance than does the best state-of-the-art ML approach, Random Forest:
a mean absolute error (MAE) of 0.038 eV/atom compared to 0.072 eV/atom on the OQMD-SC dataset.
IRNet also performed significantly better than both the plain network and SRNet.
The use of individual residual learning (IRNet) led to faster convergence compared to the existing approach of residual learning in SRNet,
while maintaining the same number of parameters.
We also evaluated IRNet performance for learning materials properties with 145 composition-based physical attributes in two other datasets: 
OQMD-C (\num{341443} data points) and MP-C (\num{83989})~\cite{jain2013the}.
IRNet significantly outperformed the plain network and the traditional ML approach on the new prediction tasks; 
the deeper models performing better in case of larger dataset (OQMD-C).
We performed a combinatorial search for materials discovery using the proposed models.
The models were trained on \num{32111} entries in OQMD-SC-ICSD dataset.
The evaluation was performed by searching for stable materials with specific crystal structures.
The proposed model provided significantly more accurate predictions compared to the traditional ML approach (Random Forest).

\section{Background}
\label{sec:bg}

\subsection{Property Prediction}
The prediction of chemical properties from material crystal structure and composition is strongly related to the discovery of new materials.
One important material property is formation enthalpy:
the change in energy when one mole of a substance in 
the standard state (1 atm of pressure and 298.15~K) is formed from its pure elements under the same
conditions~\cite{oxtoby2015principles}. In other words, it is the energy released when forming a material (chemical compound) from the constituent elements.
By knowing the formation enthalpy, one can know whether the material is stable and thus feasible to 
experimentally synthesize in laboratory.
The more negative the formation enthalpy, the more stable the compound.
Materials properties also contain various other properties~\cite{kirklin2015open, jain2013the}.

\subsection{Materials Representation}

Most ML approaches require manual feature engineering and a representation 
that incorporates domain knowledge into model inputs.
They thus take composition-based physical attributes and/or crystal structure as the input.
Recently, Ward et al.~\cite{ward2016general} presented a ML framework for formation energy prediction that used an input vector with 145 features computed from composition; stoichiometric attributes, elemental property statistics, electronic structure attributes, and ionic compound attributes. 
We leverage this approach to compute the 145 physical attributes used in our datasets.

The crystal structure of a material is defined by the shape of the unit cell and associated atom positions, which together define the repeat pattern of the atomic structures that form the material.
It is possible to represent the unit cell shape and atom positions as a vector of $3 + 3N$ features (where $N$ is the number of atoms), but this representation is not suitable for ML.
The atomic coordinates are not unique---rotating or translating the coordinate system does not change the material---and they do not readily reflect important features of the material (e.g., bond lengths).
Many crystal structure representations, such as ``bag of bonds''~\cite{Hansen2015} and histograms of bond distances~\cite{schutt2014represent}, have been developed to address this problem. 
We use the representation developed by Ward et al.~\cite{ward2017including}, which
uses 126 features derived from the Voronoi tessellation of a material.
The Voronoi tessellation of a crystal structure provides a clear description of the local environment of each atom, which is used to compute features such as the difference in elemental properties (e.g., molar mass) between an atom and its neighbor~\cite{ward2017including}.



\section{Design}
\label{sec:meth}


We next describe how we build deep residual regression models, composed of multiple fully connected layers, 
for data mining problems with numerical vectors as inputs.
We first introduce a plain network without any residual learning. 
Next, we build a stacked residual network by introducing shortcut connections for residual learning after each of a number of stacks, each composed of one or more layers with the same configuration. 
Finally, we introduce our novel individual residual learning approach, in which shortcut connections are used after every layer.
We use the plain network and stacked networks later as baseline models for comparison against the individual residual network.


\begin{figure}
\centering
\includegraphics[width=.5\textwidth,keepaspectratio=true,trim=0.4in 0 2in 0in,clip]{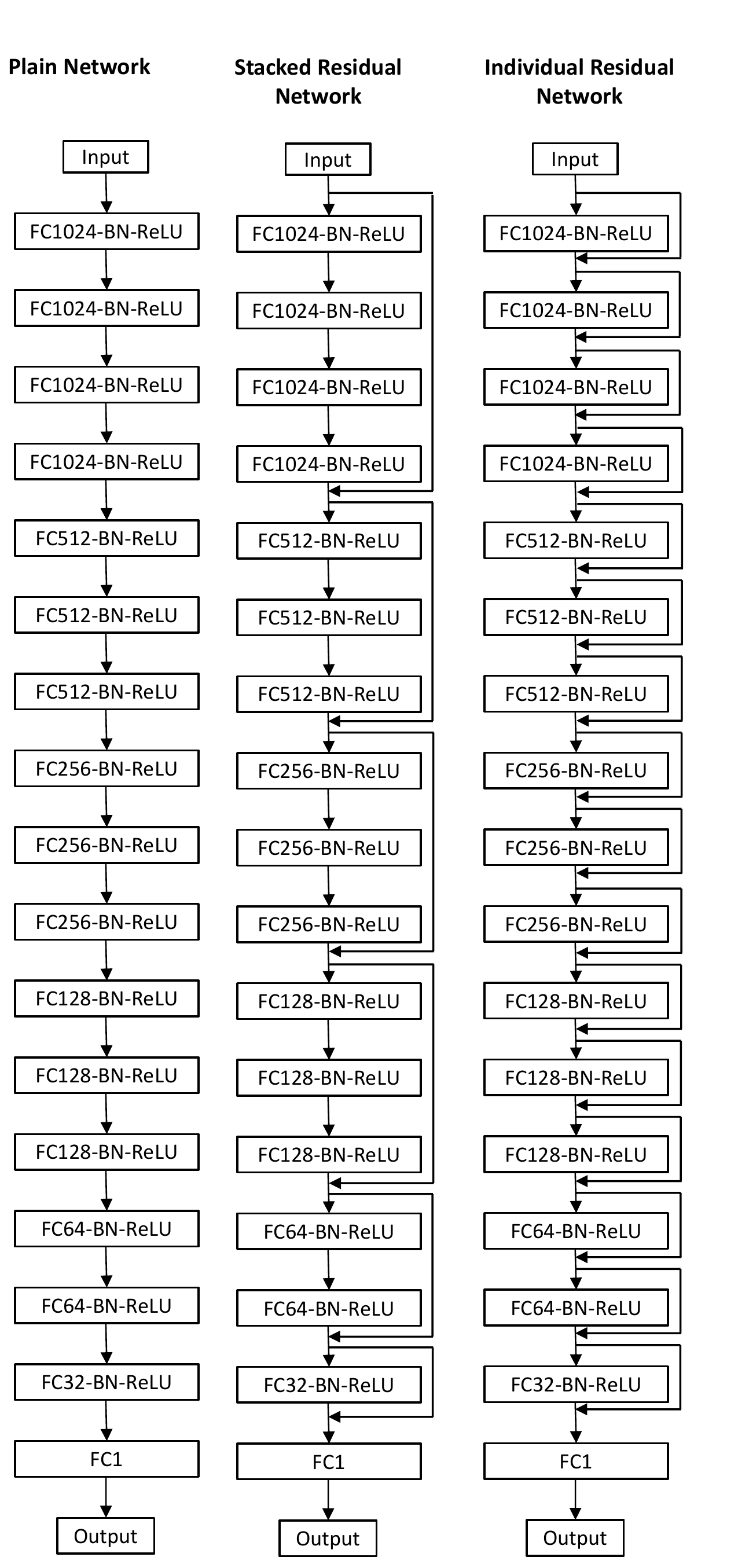} 
\caption{Three types of 17-layer networks. 
Each ``layer'' is a fully connected neural network layer with size as described in Table~\ref{tab:model_archs};
all but the last are followed by batch normalization and ReLU.
A \emph{plain network} simply connects the output of each layer to the input of the next.
A \emph{stacked residual network} (SRNet) places a shortcut connection after groups of layers called stacks. 
An \emph{individual residual network} (IRNet) places a shortcut connection after every layer.
}
\label{fig:model_archs}
\end{figure}

\begin{table*}[!tbh]
\centering
\caption{Detailed configurations for different depths of network architecture. 
The notation $[...]$ represents a stack of model components, comprising a single  
(FC: fully connected layer, BN: batch normalization, Re: ReLU activation function) sequence
in the case of IRNet and multiple such sequences in the case of SRNet.
Each such stack is 
followed by a shortcut connection. 
}
\label{tab:model_archs}
\begin{tabular}{|c|c|c|c|c|c|c|}
\hline

\textbf{Output}&\textbf{17-layer SRNet} &\textbf{17-layer IRNet} &\textbf{24-layer SRNet} &\textbf{24-layer IRNet} &\textbf{48-layer SRNet} &\textbf{48-layer IRNet} \\ \hline


1024 & [FC1024-BN-Re x 4] & [FC1024-BN-Re] x 4 & [FC1024-BN-Re x 4] & [FC1024-BN-Re] x 4 & [FC1024-BN-Re x 4] x 2 &[FC1024-BN-Re] x 8 \\ \hline


512 & [FC512-BN-Re x 3] & [FC512-BN-Re] x 3 & [FC512-BN-Re x 4] & [FC512-BN-Re] x 4 & [FC512-BN-Re x 4] x 2 &[FC512-BN-Re] x 8 \\ \hline


256 & [FC256-BN-Re x 3] & [FC256-BN-Re] x 3 &[FC256-BN-Re x 4] &[FC256-BN-Re] x 4 & [FC256-BN-Re x 4] x 2 &[FC1024-BN-Re] x 8\\ \hline


128 & [FC128-BN-Re x 3] & [FC128-BN-Re] x 3 &[FC128-BN-Re x 4] &[FC128-BN-Re] x 4 & [FC128-BN-Re x 4] x 2 &[FC128-BN-Re] x 8 \\ \hline

64 &  [FC64-BN-Re x 2] & [FC64-BN-Re] x 2 & [FC64-BN-Re x 3] & [FC64-BN-Re] x 3 & [FC64-BN-Re x 4] x 2 &[FC64-BN-Re] x 8 \\ \hline


32 & [FC32-BN-Re] & [FC32-BN-Re] & [FC32-BN-Re x 2] & [FC32-BN-Re] x 2 &[FC32-BN-Re x 4] & [FC32-BN-Re] x 4 \\ \hline
16 &  &  & [FC16-BN-Re x 2] & [FC16-BN-Re] x 2 & [FC16-BN-Re x 3] & [FC16-BN-Re] x 3 \\ \hline

1 & \multicolumn{6}{c|}{FC1}\\ \hline

\end{tabular}
\end{table*}

\subsection{Plain Network}
The model architecture is formed by putting together a series of stacks,
each composed of one or more sequences of three basic components with the same configuration. 
Since the input is a numerical vector, the model uses a fully connected layer as the initial layer in each sequence. 
Next, to reduce the internal covariance drift for proper gradient flow during back propagation for faster convergence, a batch normalization layer is placed after the fully connected layer~\cite{ioffe2015batch}. 
Finally, ReLU~\cite{nair2010rectified} is used as the activation function after the batch normalization.

The simplest instantiation of this architecture adds no shortcut connections
and thus learns simply the approximate mapping from input to output.
We refer to this network as a \emph{plain network}. 

\subsection{Stacked Residual Learning}
Deep neural networks suffer from the vanishing or exploding gradient problem~\cite{bengio1994learning,glorot2010understanding}, which hampers convergence, 
and also from
the degradation problem: as network depth increases, accuracy becomes saturated and then degrades rapidly. 
One approach to dealing with these issues is to use shortcut connections for residual learning~\cite{he2016deep, huang2017densely, srivastava2015training}.

Here, we introduce the idea of residual learning to deep regression networks composed of fully connected layers.
In a fully connected network the number of parameters is directly proportional to the product of the number of inputs and the number of output units.
The gated connection approach from the highway network and the use of all previous inputs from DenseNet~\cite{huang2017densely} would result in a huge increase in model parameters that would not fit in GPU memory.
Hence, for a fully connected deep neural network, the residual learning from He et al.~\cite{he2016deep} is the most elegant approach.

We use stacks of consecutive layers with the same configuration,
with the first stack composed of four sequence of layers
and the final stack of two sequences.
Instead of directly fitting the underlying mapping, the stacked layers explicitly learn the residual mapping. 
If the underlying mapping is denoted by $H(\vec{x})$, 
the stacked layers fit the residual mapping of $F(\vec{x}) = H(\vec{x})-\vec{x}$.
If the input and output of a stack have the same dimensions, 
they can be added by using a shortcut connection for residual learning.
If the output of a layer, $F(\vec{x})$, has a different dimension than the input $\vec{x}$, 
we perform a linear projection $W_s$ to match the dimensions before adding: 
\begin{equation}
\vec{y} = F(\vec{x}) + W_s\vec{x},
\end{equation}
where $\vec{x}$ and $F(\vec{x})$ are the input and output to the stack of layers, respectively. $W_s$ acts as 
a dimension reduction agent.
We refer to such a network with shortcut connections across each stack as a \emph{ stacked residual network} (SRNet). 

\subsection{Individual Residual Learning}
He et al.~\cite{he2016deep} introduced the idea of using shortcut connections after a stack composed of multiple convolutional layers.
The latest Inception-ResNet~\cite{szegedy2017inception} architecture for image classification follows a similar approach,
with shortcut connections used between stack of Inception-ResNet blocks, 
where each block is composed of multiple convolutional layers followed by $1\times1$ convolutional filters for dimension matching.
In our case, the stacks are composed of up to four sequences, with each sequence containing a fully connected layer, a batch normalization, and ReLU.
Our stacks are comparably more complex and highly non linear when compared to those used in CNN models for image classification.
Also, learning the residual regression mapping from input to output vector is comparatively harder than the residual learning for classification task; the activations and gradients can vanish within the stacks.

To solve this issue, we introduce a novel technique of individual residual learning for sequences containing a fully connected layer with batch normalization and non linear activation.
We place a shortcut connection after every sequence, so that each sequence needs only to learn the residual mapping between its input and output. This innovation has the effect of making the regression learning task easy.
As each ``stack'' now comprises a single sequence,
shortcut connections across each sequence provide a smooth flow of gradients between layers. 
We refer to such a deep regression network with individual residual learning capability as an \emph{individual residual network} (IRNet).

The detailed architectures for networks with different depths are illustrated in Figure~\ref{fig:model_archs} and Table~\ref{tab:model_archs}.
There are several deep network design techniques based on advanced branching techniques such as Inception~\cite{szegedy2015going, szegedy2017inception} and ResNext~\cite{xie2017aggregated}, but here our goal is to design a general purpose deep regression network framework rather than optimizing for a specific prediction task. 
We will explore branching techniques in future work.


\section{Empirical Evaluation}
\label{sec:exp} 
We now present a detailed analysis of the design and evaluation of our deep regression networks with residual learning.
We proceed in three stages.
First, we present our evaluation of the proposed deep regression model (IRNet) for the \emph{design problem} and compare its performance with the plain network, SRNet, and traditional ML approaches when applied to the OQMD-SC dataset.  
Next, we evaluate the proposed model architecture by learning materials properties from physical attributes for compounds in the OQMD-C and MP-C datasets. Finally, we perform a combinatorial search for materials discovery by training on the OQMD-SC-ICSD dataset. 
Before presenting our evaluation, we discuss the experimental settings and datasets that we use in this work.

\subsubsection*{Experimental Settings}
We implement the deep learning models with Python and TensorFlow~\cite{abadi2016tensorflow}.
We performed extensive architecture search and hyperparameter tuning for all deep learning and other ML models used in this study.
For deep learning models, we experimented with different activation functions: sigmoid, tanh, and ReLU, both for the intermediate layers and for the final regression layer. 
We explored learning rates in [1e-1, 1e-2, 1e-3, 1e-4, 1e-5, 1e-6];
StochasticGradientDescent, MomentumOptimizer, Adam, and RMSProp optimizers;
and mini-batch sizes in [32, 64, 128].
Since we are dealing with regression output, we experimented with mean squared error and mean absolute error as the loss functions.
We found the best hyperparameters to be are Adam~\cite{kingma2014adam} as the optimizer with a mini batch size of 64, learning rate of 0.0001, mean absolute error as loss function, and ReLU as activation function, with the final regression layer having no activation function.
Rather than training the model for a specific number of epochs, we used early stopping with a patience of 200, meaning that
we stopped training when the performance did not improve in 200 epochs.
For traditional ML models, we used Scikit-learn~\cite{articlescikit} implementations
and employed mean absolute error (MAE) as loss function and error metric.


\subsubsection*{Datasets}
We used four datasets to evaluate our models:
OQMD-SC, OQMD-C, MP-C, and OQMD-SC-ICSD. 
OQMD-SC is composed of \num{435582} unique compounds (unique combination of composition and crystal structure) with their DFT-computed formation enthalpy from the Open Quantum Database (OQMD)~\cite{kirklin2015open}; this is used for the \emph{design problem}.
It is composed of 271 attributes: 125 derived to represent crystal structure using Voronoi tesselations and another 145 physical attributes derived from composition using domain knowledge, as in Ward et al.~\cite{ward2016general}.
OQMD-C is composed of \num{341443} compounds with the materials properties from OQMD as of May 2018.
MP-C is composed of \num{83989} inorganic compounds from the Materials Project database~\cite{jain2013the} with a set of materials properties as of September 2018.
OQMD-C and MP-C contain composition only (no structure information); we compute 145 physical attributes from the composition using Ward et al.'s methods~\cite{ward2016general}.
OQMD-SC-ICSD is composed of entries from the Inorganic Crystal Structure Database (ICSD)~\cite{icsd} present in OQMD-SC.
The datasets are randomly split into training and test sets in the ratio of 9:1. 

\subsection{Design Problem}
First, we analyze the impact of different design choices by evaluating the proposed models on the design problem.
The design problem involves learning to predict formation enthalpy from input vector composed of 126 attributes to represent crystal structure 
and 145 physical attributes in OQMD-SC dataset.
An extensive architecture search and hyperparameter tuning is performed to search for the best deep regression model for the design problem.

\begin{figure}[!tb]
\centering
\includegraphics[width=.99\columnwidth,keepaspectratio=true,trim=0.18in 0.2in 0.17in 0.17in,clip]{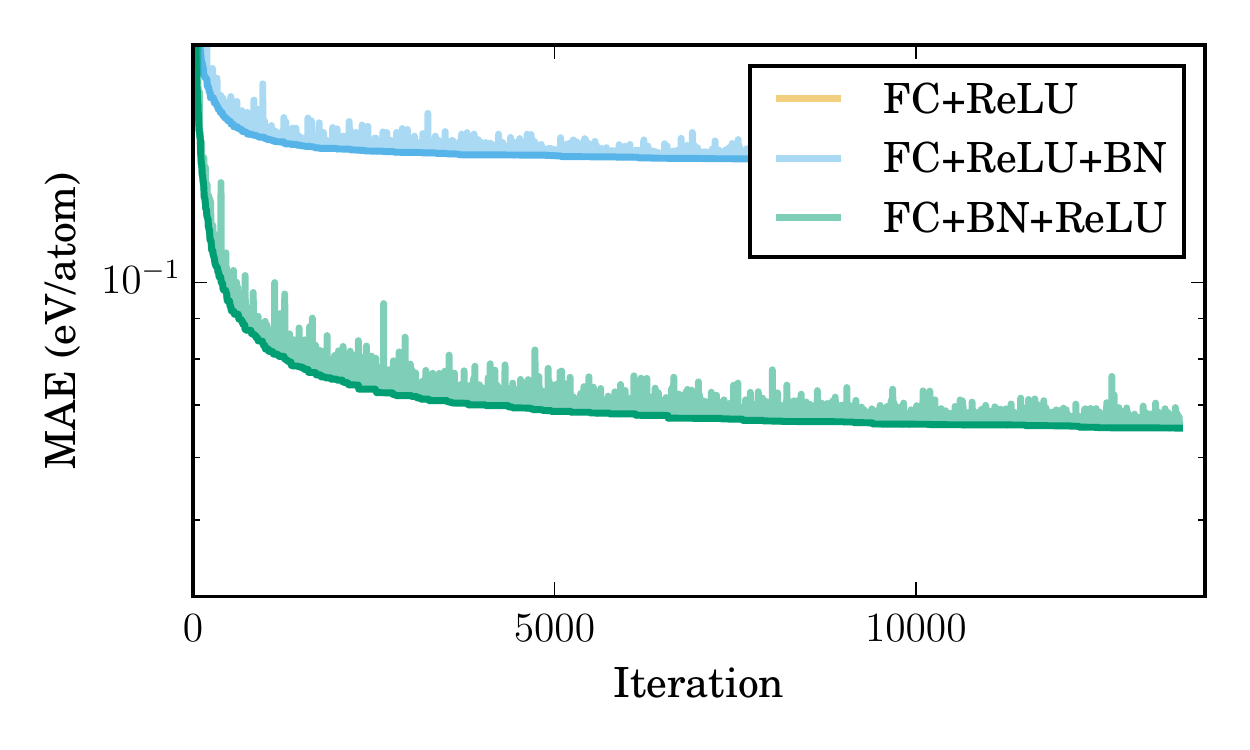}
\caption{Test error curve for various plain networks for the \emph{design problem}. 
Batch normalization before activation function (FC+BN+ReLU) improves performance significantly. }
\label{fig:impact_comp}
\end{figure}

\subsubsection{Basic Components} 
We experimented with different patterns of use of our basic components---fully connected layer, batch normalization, activation function, and dropout---within the plain network.
Use of batch normalization resulted in significant reduction in errors, as seen in Figure~\ref{fig:impact_comp}.
Batch normalization can be used either before (FC+BN+ReLU) or after the activation function (FC+ReLU+BN). 
For our regression problem, using batch normalization before ReLU (FC+BN+ReLU) worked better; 
the original work also used it before the activation function for image classification problem~\cite{ioffe2015batch}.
Since ReLU truncates all negative activations to zero, applying batch normalization on ReLU outputs leads to changes in the activation distribution; since the regression output is dependent on all activations, 
batch normalization after ReLU leads to higher oscillations and poor convergence.

We also experimented with using dropouts after the first four stacks for better generalization; however, dropouts resulted in slight degradation in the performance. 
The best plain network architecture for our design problem is composed of 17 sequences containing a fully connected layer, a batch normalization and a ReLU; we refer to this as the \emph{17-layer plain network}. 
as shown in Figure~\ref{fig:model_archs}.

\begin{figure}[!tb]
\centering
\includegraphics[width=.99\columnwidth,keepaspectratio=true,trim=0.18in 0.2in 0.17in 0.17in,clip]{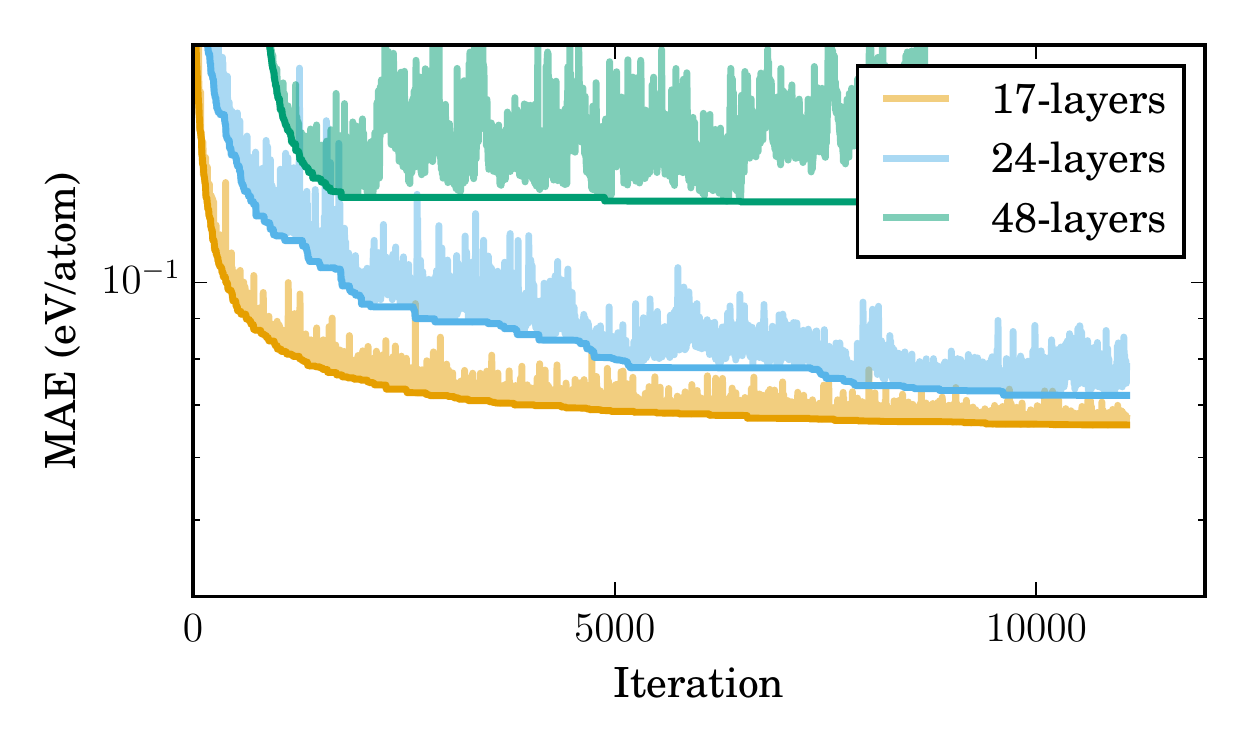}
\caption{Test error curve for deeper plain networks for the \emph{design problem}. Performance degrades with network depth, even in the presence of batch normalization.}
\label{fig:impact_deep}
\end{figure}

\begin{figure*}[tbh]
\centering

\subfigure[17-layers]{
\includegraphics[width=.32\textwidth,keepaspectratio=true,trim=0.18in 0.2in 0.15in 0.15in,clip]{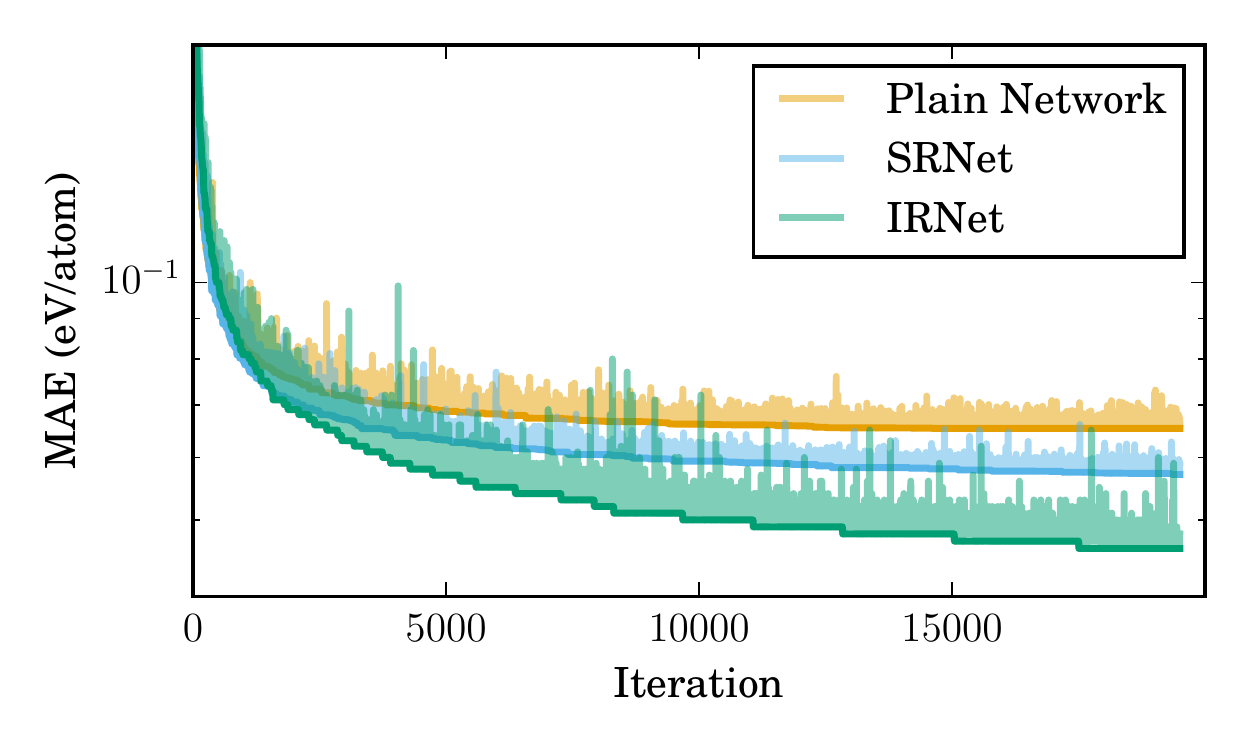}
}
\subfigure[24-layers]{
\includegraphics[width=.32\textwidth,keepaspectratio=true,trim=0.18in 0.2in 0.15in 0.15in,clip]{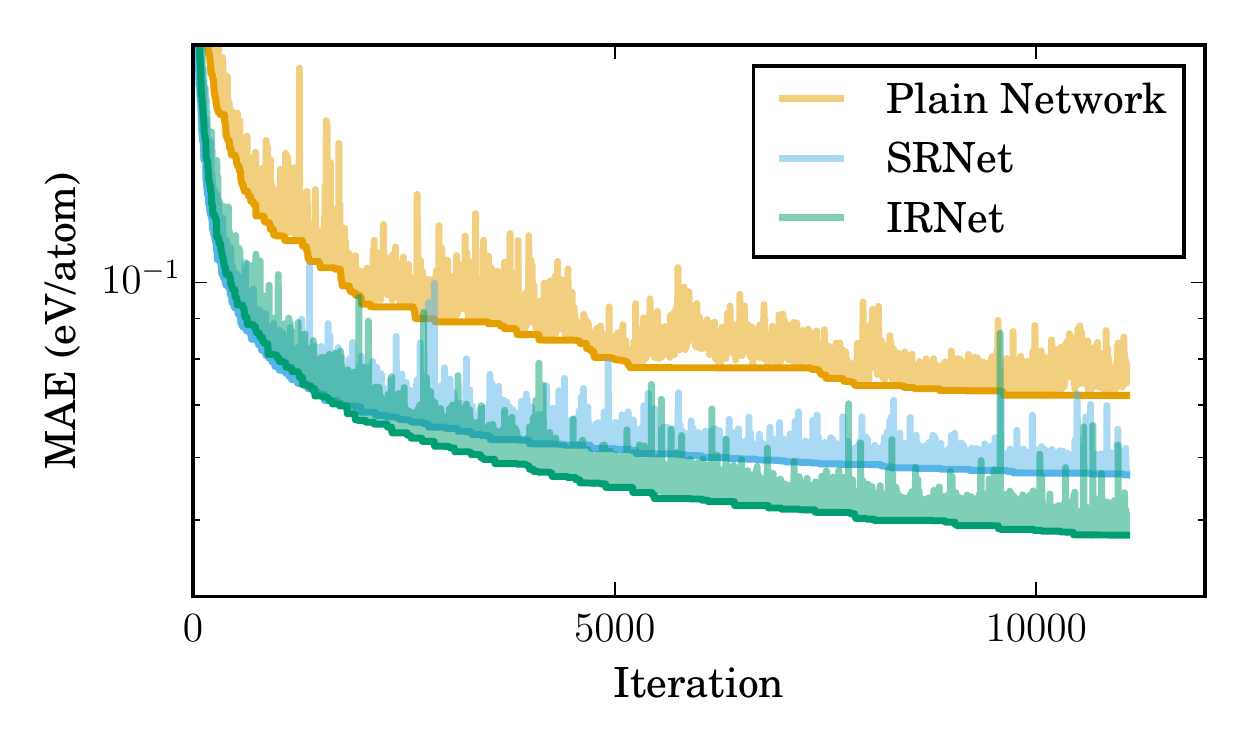}
}
\subfigure[48-layers]{
\includegraphics[width=.32\textwidth,keepaspectratio=true,trim=0.18in 0.2in 0.15in 0.15in,clip]{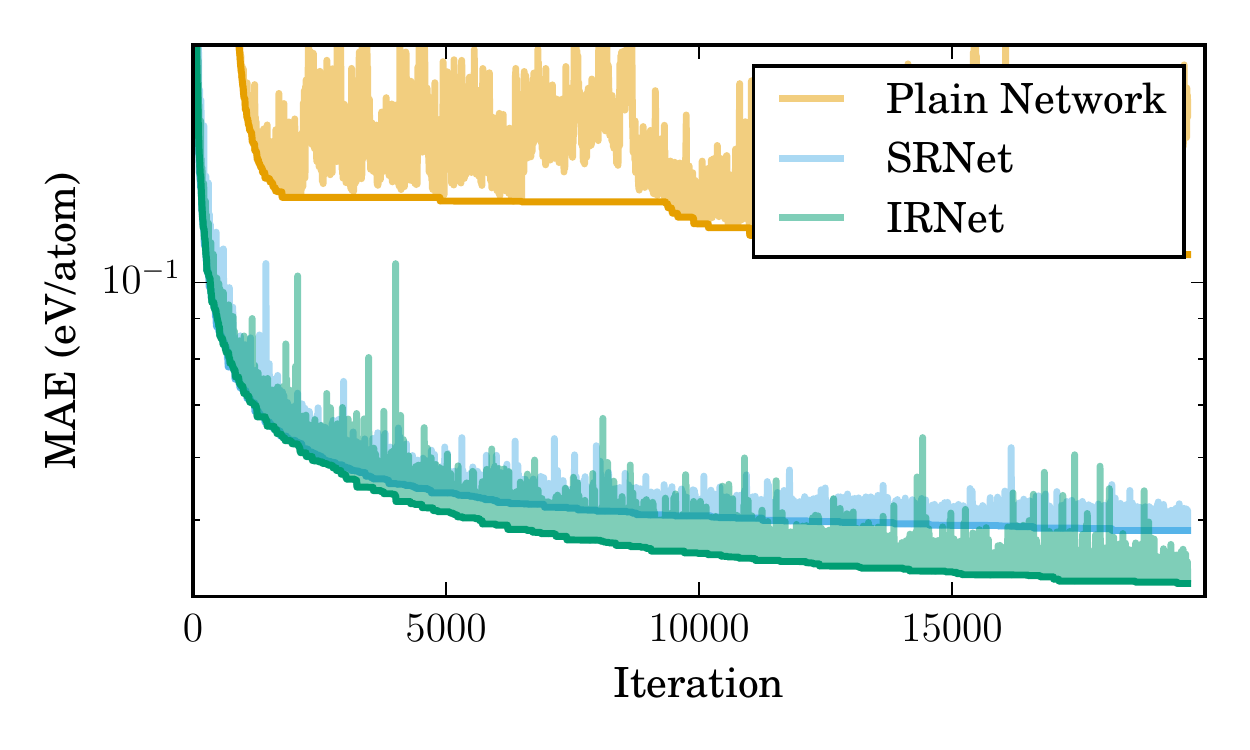}
}


\caption{Impact on residual learning for the \emph{design problem}. 
Both residual networks outperform the plain network, and the individual network outperforms the stacked network for all depths of network. We observe similar trends even in the case of training error curves for all types of networks of all depths; the IRNet converges faster than the SRNet and Plain Network for all depths.}
\label{fig:impact_res}
\end{figure*}

\subsubsection{Residual Learning}
Figure~\ref{fig:impact_deep} shows how performance can degrade with increased depth for plain networks.
This happens mainly because of the vanishing gradient problem.
To solve this issue, we introduced residual learning to create SRNet and IRNet, as discussed earlier.
We see in Table~\ref{tab:deep_results} and Figure~\ref{fig:impact_res} that the introduction of
shortcut connections to enable residual learning significantly improved model performance, 
presumably by helping with the smooth flow of gradients from output to input.
We compared the individual residual learning in IRNet with the existing approach of use of shortcut connections after stacks of multiple layers in SRNet.
The stacks are formed by putting the consecutive layers with equal number of output units in a stack.

We observe a significant benefit from the novel approach of using shortcut connections for individual residual learning in IRNet; the mean absolute error significantly decreased compared to SRNet as seen in Figure~\ref{fig:impact_res} and Table~\ref{tab:deep_results}.
Both the training and test error curves in the case of IRNet exhibits better convergence than both SRNet and plain network during the training.
We conjecture that learning the residual between the output and the input vector of the sequence is better compared to learning the  
more complex residual mapping in the case of stacked residual network in SRNet.
Also, if the identity mapping using shortcut connections are optimal, the residuals would be pushed to zero and hence, better suited for batch normalization to learn our regression output.
This illustrates the advantage of using individual residual learning for deep regression networks composed of fully connected layers for vector inputs.

\begin{table}[!tb]
\caption{Performance of deeper residual networks for the \emph{design problem}. Test errors are MAE in eV/atom. Increased depth of residual network architectures leads to improved performance for both stacked and individual residual networks. The individual residual network (IRNet) clearly outperforms the stacked residual network (SRNet), achieving significantly lower MAE.}
\label{tab:deep_results}
\begin{tabular}{|l|c|P{0.6in}|P{0.6in}|}
\hline
\textbf{Model Type}&\textbf{Plain Network} & \textbf{SRNet} &{\textbf{IRNet}} \\

\hline
17-layer & 0.0653 & 0.0551 & \textbf{0.0411} \\ \hline
24-layer & 0.0719 & 0.0546 & \textbf{0.0403} \\ \hline
48-layer & 0.1085 & 0.0471 & \textbf{0.0382} \\ \hline
\end{tabular}
\end{table}

\subsubsection{Deeper Architectures}
Next, we  experimented with deeper architectures composed of 24 and 48 sequences of layers for all types of deep regression networks: plain network, SRNet, and IRNet.
From Figure~\ref{fig:impact_deep}, we can clearly observe the performance degradation issue in plain networks that do not leverage any shortcut connections for residual network.
Figure~\ref{fig:impact_res} illustrates the trend in error curves.
Although both types of residual networks exhibit reduced test error with increased depth,
the rate of reduction for IRNet is significantly better than that for SRNet.
To prevent overfitting of such deep models with large numbers of parameters to the training dataset, 
we used early stopping with a patience of 200.
Table~\ref{tab:deep_results} shows the final MAE for all types of networks with different depths.
Our results illustrates the efficiency of using individual residual learning with deeper architectures.

\begin{table}[!tb]
\caption{Performance of Traditional ML Approaches for the \emph{design problem}. We performed extensive grid search for hyperparameter tuning for all the listed ML models. 
Test errors are MAE in  eV/atom.}
\label{tab:ml_results}
\begin{tabular}{|l|P{0.6in}|}
\hline
\textbf{ML Approach}&\textbf{Test Error} \\ \hline
AdaBoost & 0.479 \\ \hline
ElasticNet & 0.384 \\ \hline
LinearRegression & 0.261 \\ \hline
Ridge & 0.261 \\ \hline
SVR & 0.243 \\ \hline
KNeighbors & 0.154 \\ \hline
DecisionTree & 0.104 \\ \hline
Bagging & 0.078 \\ \hline
RandomForest & \textbf{0.072} \\ \hline
\end{tabular}
\end{table}

\subsubsection{Comparison with Other ML Approaches}
\begin{figure}[!tb]
\centering
\includegraphics[width=.99\columnwidth,keepaspectratio=true,trim=0.18in 0.2in 0.17in 0.17in,clip]{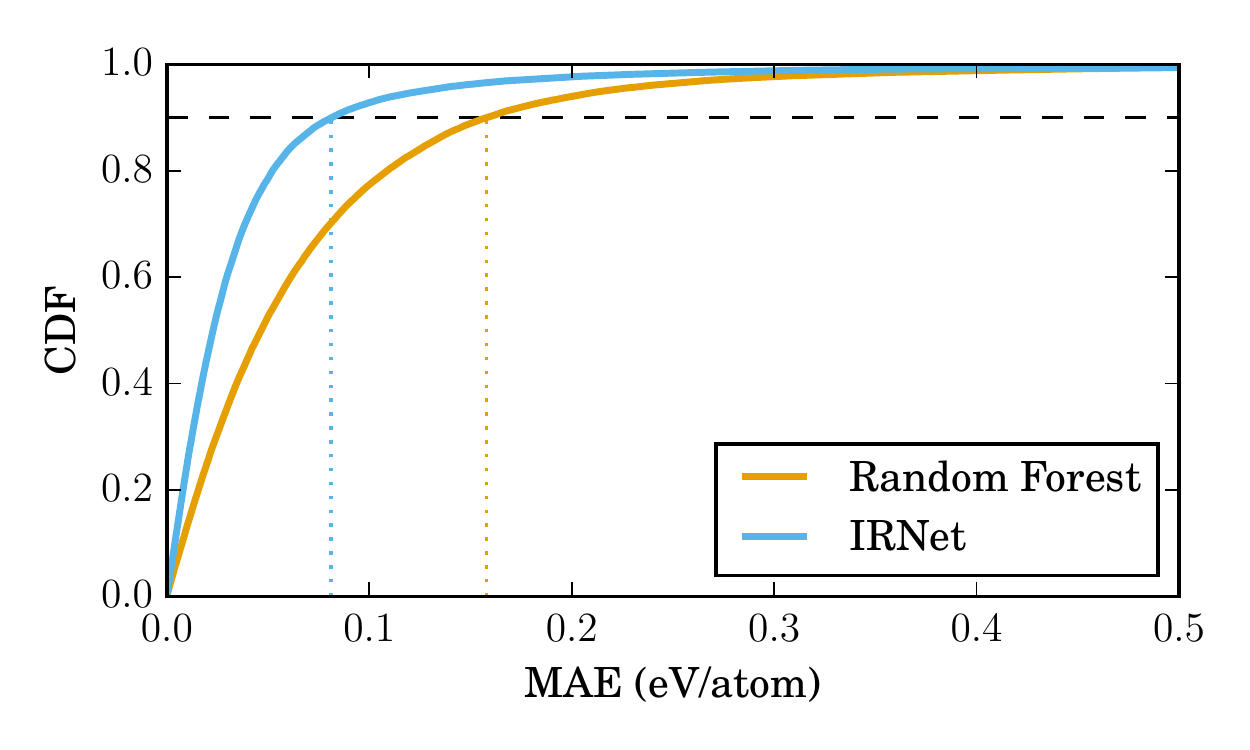}
\caption{
Cumulative distribution function (CDF) of the prediction errors for the \emph{design problem}. Deep learning (IRNet) performs significantly better than the traditional ML approach, Random Forest, achieving
a 90th percentile MAE of 0.081 eV/atom vs.\ 0.158 eV/atom for Random Forest.}
\label{fig:comp}
\end{figure}

Next, we compared the performance of the proposed deep learning model with traditional ML models: see Table~\ref{tab:ml_results}.
We performed an extensive hyperparameter search to find the best hyperparameters for all ML models.
For instance, for Random Forest model, we used a minimum sample split from [5, 10, 15, 20], number of estimators from [100,150,200], maximum features from [0.25, 0.33] and maximum depth from [10,25].
Similarly, extensive grid search for optimization of hyperparameters for other ML models are used.
Among all of the traditional ML approaches considered,
Random Forest achieved the best MAE of 0.072 eV/atom.
By comparison, the 48-layer IRNet achieved an MAE of 0.038 eV/atom,
significantly outperforming Random Forest for the design problem.
Figure~\ref{fig:comp} illustrates the comparison of the prediction errors for the test set.
Deep learning provides a more accurate and robust prediction model than does the state-of-the-art ML approach, Random Forest,
predicting the formation enthalpy of 90\% of the compounds in the test set with half the error of Random Forest.
These results demonstrate that deep learning in general, and IRNet in particular, can help construct a robust model for predicting formation enthalpy from materials crystal structure and composition.

\subsubsection{Summary of design insights}
We draw the following lessons from our experiments with building deep regression networks for learning regression output from numerical vector inputs.

\begin{enumerate}

\item \textbf{Batch Normalization} Batch normalization works better in deep regression networks if used before ReLU. Otherwise, ReLU truncates all negative values to zero, which makes learning the regression output hard. 
Dropout with batch normalization slightly worsens performance.


\item \textbf{Residual Learning} Residual learning in deep regression always performs better compared to directly learning to fit the underlying mapping from input vector to the regression output.

\item \textbf{Individual Residual Learning} Putting a shortcut connection after each sequence of layers (IRNet) works significantly better than the conventional way of putting the shortcut connection after each stack of multiple layers (SRNet).


\end{enumerate}

The presented architecture can be applied to other data mining problems with vector inputs in scientific domains; they can provide more robust and accurate predictive modeling than the existing ones based on traditional ML approach.
The same architecture can be also applied to classification problem by adding a $softmax$ activation at the last layer and using $cross$ $entropy$ as the loss function.

\subsection{Other Datasets}
We evaluated the proposed deep regression architecture on learning materials properties present in two other datasets, OQMD-C and MP-C.
OQMD-C is composed of \num{341443} samples while MP-C has \num{83989} samples; they contain the materials properties with their composition.
For comparison, we used the 17-layered plain network and ten other traditional ML approaches.
We did not perform hyperparameter tuning and architecture search for deep learning models for these tasks, to illustrate the general purpose use of the proposed deep regression model.
The deep regression networks designed for the \emph{design problem} were trained on an input vector containing 145 physical attributes derived from composition; they were trained from scratch using random weights initialization.
For the traditional ML models, we performed an extensive grid search for hyperparameter optimization as in the previous case for the design problem.

We can observe three things from the results in Table~\ref{tab:other_data}.
First, the deep learning network almost always outperforms the traditional ML approaches.
Second, the proposed network with individual residual learning performs better than the plain network in all cases.
Third, deeper networks worked better in case of OQMD-C while they did not help in case of MP-C, suggesting that
deeper networks work better when the dataset size is larger (OQMD-C vs MP-C). This agrees with the fact that deep neural networks perform better with big data.
The results demonstrate that although the proposed model was originally designed for a different \emph{design problem},
they almost always outperform the plain network and the traditional ML approaches used by domain scientists.
We also experimented with SRNet from design problem for these prediction problems, SRNet performed better than the plain network but worse than the IRNet, similar to the results for the design problem.
This illustrate that IRNet can serve as a general purpose deep learning model for different predictive modeling tasks where we need to learn the regression output from an input vector composed of materials composition and/or crystal structures.

\begin{table*}[!tbh]
\caption{Performance on OQMD-C and MP-C datasets of our DNN models vs.\ 10 traditional ML approaches for regression problems: Linear Regression, Lasso, Ridge, Decision Tree, Adaboost, KNeighbors, ElasticNet, SGD Regression, Random Forest and Support Vector, with extensive grid search used to tune hyperparameters for each. Test errors are MAE in eV/atom. 
}
\label{tab:other_data}
\begin{tabular}{|l| l |c|c|c|c|}
\hline
\textbf{Dataset} &\textbf{Property}&\textbf{Best of 10 ML} &\textbf{17-layer Plain Network} & \textbf{17-layer IRNet} &{\textbf{48-layer IRNet}} \\

\hline \cline{1-6}
\multirow{4}{*}{OQMD-C} & Formation Enthalpy & 0.077 & 0.072 & 0.054 & \textbf{0.048} \\
\cline{2-6}
 & Bandgap & \textbf{0.047} & 0.052 & 0.051 & \textbf{0.047} \\ \cline{2-6}
 & Energy\_per\_atom & 0.1139 & 0.0939 & \textbf{0.0696} & - \\ \cline{2-6}
 & Volume\_pa & 0.473 & 0.0.483 & 0.415 & \textbf{0.394} \\ \hline\cline{1-6}

\multirow{7}{*}{MP-C} & Bandgap & 0.4788 & 0.396 & \textbf{0.363} & 0.364 \\ \cline{2-6}
& Density & 0.5052  & 0.401 & \textbf{0.348} & 0.386 \\ \cline{2-6}
& Energy\_above\_hull & 0.1184 & 0.098 & \textbf{0.091} & 0.0944 \\ \cline{2-6}
& Energy\_per\_atom & 0.2999 & 0.175 & \textbf{0.143} & - \\ \cline{2-6}
& Total\_magnetization & 3.232 & 3.0897 & \textbf{3.005} & -\\ \cline{2-6}
& Volume & 225.671 & 219.439 & \textbf{215.037} & - \\ \hline\cline{1-6}
\end{tabular}
\end{table*}

\subsection{Application for Materials Discovery}

Since the proposed model achieved a significant reduction in prediction error for formation enthalpy compared to state-of-the-art approach, it can be applied for high throughput materials discovery.
To test the ability of the proposed method to identify new materials, we emulated a common approach in computational materials science, namely combinatorial search .
A combinatorial search involves first enumerating all possible combinations of different elements on a specific crystal structure prototype, 
and then evaluating the stability of each resultant structure with DFT to find which are stable.
We performed a combinatorial search using the evaluation settings based
on the combinatorial search analysis from~\cite{ward2017including}.
OQMD-SC-ICSD, used as a training set by Ward et al.~\cite{ward2017including}, 
comprises \num{32111} entries in OQMD-SC that correspond to known, experimentally-synthesized materials in ICSD~\cite{icsd}.
The proposed IRNet is trained using the OQMD-SC-ICSD dataset and evaluated by
predicting the formation enthalpy (stability) of materials with crystal structures from three different, commonly occurring crystal structure types: B2, L1$_0$, and orthorhombically-distorted perovskite.
These three structure types were chosen to sample structures with different kinds of bonding environments and that are stable with different types of chemistry (e.g., metals vs.\ oxides).

\begin{table}[!tb]
\caption{Performance from combinatorial search. 
Our 17-layer IRNet, when trained on OQMD-SC-ICSD, predicts formation enthalpy (stability) more accurately than Random Forest for all three types of crystal structures considered. 
}
\label{tab:comb_search}
\begin{tabular}{|l|c|c|}
\hline
\textbf{Crystal} &\textbf{Random Forest} & \textbf{17-layers IRNet } \\
\textbf{Structure} & \textbf{MAE (eV/atom)} & \textbf{MAE (eV/atom)} \\
\hline
B2 & 0.5114 & \textbf{0.4780}  \\ \hline
L$1_0$ & 0.4793 & \textbf{0.4419} \\ \hline
Perovskite & 0.6166 & \textbf{0.3693} \\ \hline
\end{tabular}
\end{table}

We show in Table~\ref{tab:comb_search} the deep learning model's prediction error for each type of crystal structures. To compare the performance of our deep learning model, we also trained a Random Forest model (the best traditional ML approach from previous analysis) on OQMD-SC-ICSD, with extensive hyperparameter search.
Our results demonstrate that our models perform better on the evaluation candidates than does the Random Forest model.
Although we do not repeat the entire combinatorial search workflow here with the proposed models, more accurate predictions on the discoveries from Ward et al.~\cite{ward2017including} suggest that the proposed IRNet model can improve the quality and robustness of the combinatorial search workflow.
Despite a small training data size, the IRNet model provides a more robust method for performing combinatorial search for high-throughput materials discovery.


\section{Conclusions and Future Work}
\label{sec:conc}
In this paper, we studied and proposed the design principles for building deep regression networks composed of fully connected layers for data mining problems with numerical vector input.
We introduced the use of residual learning in deep regression network; we proposed a deep regression network (IRNet) that leveraged individual residual learning in each layer.
The proposed IRNet outperformed the plain network (without residual learning) and traditional machine learning approaches in learning different materials properties from different size of datasets and input vector.
For the \emph{design problem} of predicting formation enthalpy from crystal structures and composition, the proposed IRNet significantly reduced the MAE from 0.072 eV/atom to 0.038 eV/atom.
We were able to converge the deep regression networks with up to 48 layers, performance increasing with greater depth.
Since IRNet kept improving performance with increased depth,
we plan to explore deeper IRNet architectures to study their impact on model performance and convergence, 
and to apply the resulting networks to data mining problems from other scientific domains.
It will also be interesting to see how this model performs on experimental datasets using transfer learning from larger simulation datasets.
The proposed deep learning model and design insights gained from this work can be used in building predictive models for other applications with vector inputs.
The code repository is available at \url{https://github.com/dipendra009/IRNet}; we also plan to make the models described in this work available via DLHub~\cite{chard19dlhub}.

\begin{acks}
This work was performed under the following financial assistance award 70NANB19H005 from U.S. Department of Commerce, National Institute of 
Standards and Technology as part of the Center for Hierarchical Materials Design (CHiMaD). Partial support is also acknowledged from DOE awards DE-SC0014330, DE-SC0019358.
\end{acks}
\bibliographystyle{ACM-Reference-Format}
\bibliography{crystal_net}

\end{document}